\documentclass[a4paper, 12pt]{article}
\pdfoutput=1

\usepackage[latin1]{inputenc}

\usepackage{amssymb}
\usepackage{amsmath}

\usepackage{natbib}

\usepackage{graphics}
\usepackage{graphicx}
\usepackage{xspace}

\usepackage{bm,lineno}
\usepackage{authblk}

\let\proglang=\textsf
\def\RR{\textsf{R}\xspace}

\let\pkg=\strong

\newcommand\code{\bgroup\@codex}
\def\@codex#1{\small {\normalfont\ttfamily\hyphenchar\font=45 #1}\egroup}

\title{Markov Chain Monte Carlo with the Integrated Nested Laplace Approximation}

\author[1]{Virgilio G\'omez-Rubio\thanks{Virgilio.Gomez@uclm.es}}
\author[2]{H\aa{}vard Rue\thanks{haavard.rue@kaust.edu.sa}}

\affil[1]{
Department of Mathematics, School of Industrial Engineering\\
Universidad de Castilla-La Mancha\\
Avda Espa\~na s/n, 02071 Albacete, Spain}
\affil[2]{
CEMSE Division\\
King Abdullah University of Science and Technology\\
Thuwal 23955-6900, Saudi Arabia}

\begin{document}

\maketitle

\begin{abstract}

    The Integrated Nested Laplace Approximation (INLA) has established
    itself as a widely used method for approximate inference on
    Bayesian hierarchical models which can be represented as a latent
    Gaussian model (LGM). INLA is based on producing an accurate
    approximation to the posterior marginal distributions of the
    parameters in the model and some other quantities of interest by
    using repeated approximations to intermediate distributions and
    integrals that appear in the computation of the posterior
    marginals.

    INLA focuses on models whose latent effects are a Gaussian Markov
    random field (GMRF). For this reason, we have explored alternative
    ways of expanding the number of possible models that can be fitted
    using the INLA methodology. In this paper, we present a novel
    approach that combines INLA and Markov chain Monte Carlo (MCMC).
    The aim is to consider a wider range of models that cannot be
    fitted with INLA unless some of the parameters of the model have
    been fixed. Hence, conditioning on these parameters the model
    could be fitted with the \pkg{R-INLA} package. We show how new
    values of these parameters can be drawn from their posterior by
    using conditional models fitted with INLA and standard MCMC
    algorithms, such as Metropolis-Hastings. Hence, this will extend
    the use of INLA to fit models that can be expressed as a
    conditional LGM. Also, this new approach can be used to build
    simpler MCMC samplers for complex models as it allows sampling
    only on a limited number parameters in the model.

    We will demonstrate how our approach can extend the class of
    models that could benefit from INLA, and how the \pkg{R-INLA}
    package will ease its implementation. We will go through simple
    examples of this new approach before we discuss more advanced
    problems with datasets taken from relevant literature.

\textbf{Keywords:} Bayesian Lasso, INLA, MCMC, Missing Values, Spatial Models

\end{abstract}

\section{Introduction}

Bayesian inference for complex hierarchical models has almost entirely
relied upon computational methods, such as Markov chain Monte Carlo
\citep[MCMC,][]{Gilksetal:1996}. \citet{isi:000264374200002} propose a
new paradigm for Bayesian inference on hierarchical models that can be
represented as latent Gaussian models (LGMs), that focuses on
approximating marginal distributions for the parameters in the model.
This new approach, the Integrated Nested Laplace Approximation (INLA,
henceforth), uses several approximations to the conditional
distributions that appear in the integrals needed to obtain the
marginal distributions. See Section \ref{sec:INLA} for details.

INLA is implemented as an R package, called \pkg{R-INLA}, that allows
us to fit complex models often in a matter of seconds. Hence, this is
often much faster than fitting the same model using MCMC methods.
Fitting models using INLA is restricted, in practice, to the classes
of models implemented in the \pkg{R-INLA} package. Several authors
have provided ways of fitting other models with INLA by fixing some of
the parameters in the model so that conditional models are fitted with
\pkg{R-INLA}. We have included a brief summary below.

\citet{Lietal:2012} provide an early application of the idea of
fitting conditional models on some of the model parameters with
\pkg{R-INLA}. They developed this idea for a very specific example on
spatiotemporal models in which some of the models parameters are fixed
at their maximum likelihood estimates, which are then plugged-in the
overall model, thus ignoring the uncertainty about these parameters
but greatly reducing the dimensionality of the model. However, they do
not tackle the problem of fitting the complete model to make inference
on all the parameters in the model.

\citet{Bivandetal:2014,Bivandetal:2015} propose an approach to extend
the type of models that can be fitted with \pkg{R-INLA} and apply
their ideas to fit some spatial models. They note how some models can
be fitted after conditioning on one or several parameters in the
model. For each of these conditional models \pkg{R-INLA} reports the
marginal likelihood, which can be combined with a set of priors for
the parameters to obtain their posterior distribution. For the
remainder of the parameters, their posterior marginal distribution can
be obtained by Bayesian model averaging \citep{Hoetingetal:1999} the
family of models obtained with \pkg{R-INLA}.

Although \citet{Bivandetal:2014,Bivandetal:2015} focus on some spatial
models, their ideas can be applied in many other examples. They apply
this to estimate the posterior marginal of the spatial autocorrelation
parameter in some models, and this parameter is known to be bounded,
so that computation of its marginal distribution is easy because the
support of the distribution is a bounded interval.

For the case of unbounded parameters, the previous approach can be
applied, but a previous search may be required. For example, the
(conditional) maximum log-likelihood plus the log-prior could be
maximised to obtain the mode of the posterior marginal. This will mark
the centre of an interval where the values for the parameter are taken
from and where the posterior marginal can be evaluated.

In this paper, we will propose a different approach based on Markov
chain Monte Carlo techniques. Instead of trying to obtain the
posterior marginal of the parameters we condition on, we show how to
draw samples from their posterior distribution by combining MCMC
techniques and conditioned models fitted with \pkg{R-INLA}. This
provides several advantages, as described below.

This will increase the number of models that can be fitted using INLA
and its associated \proglang{R} package \pkg{R-INLA}. In particular,
models that can be expressed as a conditional LGM could be fitted. The
implementation of MCMC algorithms will also be simplified as only the
important parameters will be sampled, while the remaining parameters
are integrated out with INLA and \pkg{R-INLA}.
\citet{HubinStorvik:2016a} have also effectively combined MCMC and
INLA for efficient variable selection and model choice.

The paper is structured as follows. The Integrated Nested Laplace
Approximation is described in Section \ref{sec:INLA}. Markov chain
Monte Carlo methods are summarised in Section \ref{sec:MCMC}. Our
proposed combination of MCMC and INLA is detailed in Section
\ref{sec:INLAMCMC}. Some simple examples are developed in Section
\ref{sec:Example} and some real applications are provided in Section
\ref{sec:Applications}. Finally, a discussion and some final remarks
are provided in Section \ref{sec:Discussion}.

\section{Integrated Nested Laplace Approximation}
\label{sec:INLA}

We will now describe the types of models that we will be considering
and how the Integrated Nested Laplace Approximation method works. We
will assume that our vector of $n$ observed data
$\mathbf{y} = (y_1,\ldots,y_n)$ are observations from a distribution
in the exponential family, with mean $\mu_i$. We will also assume that
a linear predictor on some covariates plus, possibly, other effects
can be related to mean $\mu_i$ by using an appropriate link function.
Note that this linear predictor $\eta_i$ may be made of linear terms
on some covariates plus other types of terms, such as non-linear
functions on the covariates, random effects, spatial random effects,
etc. All these terms will define some latent effects $\mathbf{x}$.

The distribution of $\mathbf{y}$ will depend on a vector of
hyperparameters $\bm\theta_1$. Because of the approximation that INLA
will use, we will also assume that the vector of latent effects
$\mathbf{x}$ will have a distribution that will depend on a vector of
hyperparameters $\bm\theta_2$. Altogether, the hyperparameters can be
represented using a single vector $\bm \theta=(\bm\theta_1, \bm\theta_2)$.

From the previous formulation, it is clear that observations are
independent given the values of the latent effects $\mathbf{x}$ and
the hyperparameters $\bm \theta$. That is, the likelihood of our model
can be written down as
\begin{equation}
    \pi(\mathbf{y}|\mathbf{x},\bm\theta) =
    \prod_{i\in\mathcal{I}} \pi(y_i|x_i,\bm\theta)
\end{equation}
Here, $i$ is indexed over a set of indices
$\mathcal{I} \subseteq \{1,\ldots,n\}$ that indicates observed
responses. Hence, if the value of $y_i$ is missing then
$i \not\in \mathcal{I}$ (but the predictive distribution $y_i$ could
be computed once the model is fitted).

Under a Bayesian framework, the aim is to compute the posterior
distribution of the model parameters and hypermeters using Bayes'
rule. This can be stated as
\begin{equation}
    \pi(\mathbf{x}, \bm{\theta}|\mathbf{y}) \propto
    \pi(\mathbf{y}|\mathbf{x},\bm\theta) \pi(\mathbf{x},\bm \theta)
\end{equation}
Here, $\pi(\mathbf{x},\bm\theta)$ is the prior distribution of the latent
effects and the vector of hyperparameters. As the latent effects
$\mathbf{x}$ have a distribution that depends on $\bm\theta_2$, it is
convenient to write this prior distribution as
$\pi(\mathbf{x},\bm\theta) = \pi(\mathbf{x}|\bm\theta) \pi(\bm\theta)$.

Altogether, the posterior distribution of the latent effects and
hyperparameters can be expressed as
\begin{eqnarray}
  \pi(\mathbf{x}, \bm{\theta}|\mathbf{y}) \propto
  \pi(\mathbf{y}|\mathbf{x},\bm\theta) \pi(\mathbf{x},\bm\theta) =
  \pi(\mathbf{y}|\mathbf{x},\bm\theta) \pi(\mathbf{x}|\bm\theta)
  \pi(\bm\theta) = \nonumber\\
  \pi(\mathbf{x}|\bm\theta) \pi(\bm\theta) \prod_{i\in\mathcal{I}} \pi(y_i|x_i,\bm\theta)
  \label{eq:posterior}
\end{eqnarray}
The joint posterior, as presented in Equation~(\ref{eq:posterior}), is
seldom available in a closed form. For this reason, several estimation
methods and approximations have been developed over the years.

Recently, \citet{isi:000264374200002} have provided approximations
based on the Laplace approximation to estimate the marginals of all
parameters and hyperparameters in the model. They develop this
approximation for the family of latent Gaussian Markov random fields
models. In this case, the vector of latent effects is a Gaussian
Markov random field (GMRF). This GMRF will have zero mean (for
simplicity) and precision matrix $Q(\bm\theta_2)$.

Assuming that the latent effects are a GMRF will let us develop
Equation~(\ref{eq:posterior}) further. In particular, the posterior
distribution of the latent effects $\mathbf{x}$ and the vector of
hyperparameters $\bm{\theta}$ can be written as
\begin{equation}\label{eq2}%
    \pi(\mathbf{x}, \bm{\theta}|\mathbf{y}) \propto
    \pi(\bm{\theta}) |\mathbf{Q}(\bm{\theta})|^{n/2}
    \exp\{-\frac{1}{2}\mathbf{x}^T \mathbf{Q}(\bm{\theta})
    \mathbf{x}+\sum_{i\in\mathcal{I}}
    \log\left(\pi(y_i|x_i, \bm{\theta})\right) \}.
\end{equation}
With INLA, the aim is not the joint posterior distribution
$\pi(\mathbf{x}, \bm{\theta}|\mathbf{y})$ but the marginal
distributions of latent effects and hyperparameters. That is,
$\pi(x_j|\mathbf{y})$ and $\pi(\theta_k|\mathbf{y})$, where indices
$j$ and $k$ will take different ranges of values depending on the
number of latent effects and hyperparameters.

Before computing these marginal distributions, INLA will obtain an
approximation to $\pi(\bm{\theta}|\bf{y})$,
$\tilde\pi(\bm{\theta}|\mathbf{y})$. This approximation will later
be used to compute an approximation to marginals
$\pi(x_j|\mathbf{y})$. Given that the marginal can be written down as
\begin{equation}
    \pi(x_j|\mathbf{y}) = \int \pi(x_j|\bm{\theta}, \mathbf{y})
    \pi(\bm{\theta}| \mathbf{y}) d\bm{\theta},
\end{equation}
\noindent
the approximation is as follows:
\begin{equation}
    \tilde\pi(x_j|\mathbf{y})=
    \sum_g \tilde\pi (x_j|\bm{\theta_g}, \mathbf{y})\times
    \tilde\pi(\bm{\theta_g}|\mathbf{y})\times \Delta_g.
\end{equation}
Here, $\tilde\pi(x_j|\bm\theta_g, \mathbf{y})$ is an approximation to
$\pi (x_j|\bm{\theta_g}, \mathbf{y})$, which can be obtained using
different methods \citep[see,][for details]{isi:000264374200002}.
$\bm{\theta_g}$ refers to an ensemble of hyperparameters, that
take values on a grid (for example), with weights $\Delta_g$.

INLA is a general approximation that can be applied to a large number
of models. An implementation for the \proglang{R} programming language
is available in the \pkg{R-INLA} package at \pkg{www.r-inla.org},
which provides simple access to model fitting. This includes a simple
interface to choose the likelihood, latent effects and priors. The
implementation provided by \pkg{R-INLA} includes the computation of
other quantities of interest. The marginal likelihood
$\pi(\mathbf{y})$ is approximated, and it can be used for model
choice. As described in \cite{isi:000264374200002}, the approximation
to the marginal likelihood provided by INLA is computed as
$$
\tilde{\pi}(\mathbf{y}) = \int \frac{\pi(\bm{\theta}, \mathbf{x},
    \mathbf{y})}
{\tilde{\pi}_{\mathrm{G}}(\mathbf{x}|\bm{\theta},\mathbf{y})}
\bigg\lvert_{\mathbf{x}=\mathbf{x^*(\bm\theta)}}d \bm{\theta}.
$$
Here,
$\pi(\bm{\theta}, \mathbf{x}, \mathbf{y}) =
\pi(\mathbf{y}|\mathbf{x}, \bm{\theta})
\pi(\mathbf{x}|\bm{\theta}) \pi(\bm{\theta})$,
$\tilde{\pi}_{\mathrm{G}}(\mathbf{x}|\bm{\theta},\mathbf{y})$ is a
Gaussian approximation to $\pi(\mathbf{x}|\bm{\theta},\mathbf{y})$
and $\mathbf{x^*(\bm\theta)}$ is the posterior mode of $\mathbf{x}$ for a
given value of $\bm{\theta}$. This approximation is reliable when
the posterior of $\bm{\theta}$ is unimodal, as it is often the
case for latent Gaussian models. Furthermore,
\cite{HubinStorvik:2016b} demonstrate that this approximation is
accurate for a wide range of models.

Other options for model choice and assessment include the Deviance
Information Criterion \citep[DIC,][]{Spiegelhalteretal:2002} and the
Conditional Predictive Ordinate \citep[CPO,][]{pettit:1990}. Other
features in the \pkg{R-INLA} package include the use of different
likelihoods in the same model, the computation of the posterior
marginal of a certain linear combination of the latent effects and
others \citep[see,][for a summary of recent additions to the
software]{Martinsetal:2013}.

\section{Markov Chain Monte Carlo}
\label{sec:MCMC}

In the previous Section we have reviewed how INLA computes an
approximation of the marginal distributions of the model parameters
and hyperparameters. Instead of focusing on an approximation to the
marginals, Markov chain Monte Carlo methods could be used to
obtain a sample from the joint posterior marginal
$\pi(\bm x, \bm \theta|\bm y)$. To simplify the notation, we will
denote the vector of latent effects and hyperparameters by
$\bm z = (\bm x, \bm \theta)$. Hence, the aim now is to estimate
$\pi(\bm z|\bm y)$ or, if we are only interested on the posterior
marginals, $\pi(z_i|\bm y)$.

Several methods to estimate or approximate the posterior distribution
have been developed over the years \citep{Gilksetal:1996}. In the case
of MCMC, the interest is in obtaining a
Markov chain whose limiting distribution is $\pi(\bm z | \bm y)$. We
will not provide a summary of MCMC methods here, and the reader is
referred to \citet{Gilksetal:1996} for a detailed description.

The values generated using MCMC are (correlated) draws from
$\pi(\bm z | \bm y)$ and, hence, can be used to estimate quantities of
interest. For example, if we are interested in marginal inference on
$z_i$, the posterior mean from the $N$ sampled values
$\left\{z_i^{(j)}\right\}_{j=1}^{N}$ can be estimated using the empirical mean
of $\left\{z^{(j)}_i\right\}_{j=1}^{N}$. Similarly, estimates of the posterior
expected value of any function on the parameters $f(\mathbf{z})$ can be found
using that 

\begin{equation}
    E[f(\mathbf{z})|\bm y] \simeq \frac{1}{N}\sum_{j = 1} ^{N} f(\mathbf{z}^{(j)})
\end{equation}

\noindent
Multivariate estimates inference can be made by using the multivariate
nature of vector $\bm z^{(j)}$. For example, the posterior covariance
between parameters $z_k$ and $z_l$ could be computed by considering
samples $\left\{(z_k^{(j))}, z_l^{(j)})\right\}_{j=1}^{N}$.

\subsection{The Metropolis-Hastings algorithm}

This algorithm was firstly proposed by \citet{Metropolisetal:1953} and
\citet{Hastings:1970}. The Markov chain is generated by proposing new
moves according to a proposal distribution $q(\cdot|\cdot)$. The new
point is accepted with probability
\begin{equation}
    \alpha = \min \left\{
      1,
      \frac{\pi(\bm z^{(j+1)}|\bm y) q(\bm z^{(j)} |\bm z^{(j+1)})}{
          \pi(\bm z^{(j)}|\bm y) q(\bm z^{(j+1)}|\bm z^{(j)})}
    \right\}
    \label{eq:MH}
\end{equation}
In the previous acceptance probability, the posterior probabilities of
the current point and the proposed new point appear as
$\pi(\bm{z^{(j)}}|\bm y)$ and $\pi(\bm{z^{(j+1)}}|\bm y)$, respectively. These
two probabilities are unknown, in principle, but using Bayes' rule
they can be rewritten as
\begin{equation}
    \pi(\bm z|\bm y) = \frac{\pi(\bm y |\bm z)\pi(\bm z)}{\pi(\bm y)}
    \label{eq:bayes}
\end{equation}
Hence, the acceptance probability $\alpha$ can be rewritten as
\begin{equation}
    \alpha = \min \left\{
      1,
      \frac{\pi(\bm y |\bm z^{(j+1)})\pi(\bm z^{(j+1)}) q(\bm z^{(j)}
          |
          \bm z^{(j+1)})}{\pi(\bm y |\bm z^{(j)})\pi(\bm z^{(j)})
          q(\bm z^{(j+1)}|\bm z^{(j)})}
    \right\}
    \label{eq:MH2}
\end{equation}
This is easier to compute as the acceptance probability depends on
known quantities, such as the likelihood $\pi(\bm y |\bm z)$, the
prior on the parameters $\pi(\bm z)$ and the probabilities of the
proposal distribution. Note that the term $\pi(\bm y)$ that appears in
Equation~(\ref{eq:bayes}) is unknown but that it cancels out as it
appears both in the numerator and denominator.

In Equation~(\ref{eq:MH2}) we have described the move to sample from
the joint ensemble of model parameters. However, this can be applied
to individual paramaters one at a time, so that acceptance
probabilities will be
\begin{equation}
    \alpha = \min \left\{
      1,
      \frac{\pi(\bm y |z_i^{(j+1)})\pi(z_i^{(j+1)}) q(z_i^{(j)} |
          z_i^{(j+1)})}{\pi(\bm y |z_i^{(j)})\pi(z_i^{(j)})
          q(z_i^{(j+1)}| z_i^{(j)})}
    \right\}
    \label{eq:MHsingle}
\end{equation}

\section{INLA within MCMC}
\label{sec:INLAMCMC}

In this Section, we will describe how INLA and MCMC can be combined to
fit complex Bayesian hierarchical models. In principle, we will assume
that the model cannot be fitted with \pkg{R-INLA} unless some of the
parameters or hyperparameters in the model are fixed. This set of
parameters is denoted by $\bm z_c$ so that the full ensemble of
parameters and hyperparameters is $\bm z = (\bm z_c, \bm z_{-c})$.
Here $\bm z_{-c}$ is used to denote all the parameters in $\bm z$ that
are not in $\bm z_c$. Our assumptions are that the posterior
distribution of $\bm z$ can be split as
\begin{equation}
    \pi(\bm z |\bm y) \propto
    \pi(\bm y|\bm z_{-c}) \pi(\bm z_{-c}|\bm z_c) \pi(\bm z_c)
\end{equation}
and that $\pi(\bm y|\bm z_{-c}) \pi(\bm z_{-c}|\bm z_c)$ is a latent
Gaussian model suitable for INLA. This means that conditional models
(on $\bm z_c$) can still be fitted with \pkg{R-INLA}, i.e., we can
obtain marginals of the parameters in $\bm z_{-c}$ given $\bm z_c$.
The conditional posterior marginals for the $k$-th element in vector
$\bm z_{-c}$ will be denoted by $\pi(z_{-c,k} |\bm z_c, \bm y)$. Also, the
conditional marginal likelihood $\pi(\bm y | \bm z_c)$ can be easily
computed with \pkg{R-INLA}.

\subsection{Metropolis-Hastings with INLA}

We will now discuss how to implement the Metropolis-Hastings algorithm
to estimate the posterior marginal of $\bm z_c$. Note that this is a
multivariate distribution and that we will use block updating in the
Metropolis-Hastings algorithm. Say that we start from an initial point
$\bm z_c^{(0)}$ then we can use the Metropolis-Hastings algorithm to
obtain a sample from the posterior of $\bm z_c$.

We will draw a new proposal value for $\bm z_c$, $\bm z_c^{(1)}$,
using the proposal distribution $q(\cdot|\cdot)$. The acceptance
probability, shown in Equation~(\ref{eq:MH2}), becomes now:
\begin{equation}
    \alpha = \min \left\{
      1,
      \frac{\pi(\bm y |\bm z_c^{(j+1)})\pi(\bm z_c^{(j+1)})
          q(\bm z_c^{(j)} |\bm z_c^{(j+1)})}{\pi(\bm y |
          \bm z_c^{(j)})\pi(\bm z_c^{(j)})
          q(\bm z_c^{(j+1)}|\bm z_c^{(j)})}
    \right\}
    \label{eq:MH3}
\end{equation}
Note that $\pi(\bm y |\bm z_c^{(j)})$ and
$\pi(\bm y |\bm z_c^{(j+1)})$ are the conditional marginal likelihoods
on $\bm z_c^{(j)}$ and $\bm z_c^{(j+1)}$, respectively. All these
quantities can be obtained by fitting a model with \pkg{R-INLA} with
the values of $\bm z_c$ set to $\bm z_c^{(j)}$ and $\bm z_c^{(j+1)}$.
Hence, at each step of the Metropolis-Hastings algorithm only a model
conditional on the proposal needs to be fitted.

$\pi(\bm z_c^{(j)})$ and $\pi(\bm z_c^{(j+1)})$ are the priors of
$\bm z_c$ evaluated at $\bm z_c^{(j)}$ and $\bm z_c^{(j+1)}$,
respectively, and they can be easily computed as the priors are known
in the model. Values $q(\bm z_c^{(j)} |\bm z_c^{(j+1)})$ and
$q(\bm z_c^{(j+1)}|\bm z_c^{(j)})$ can also be computed as the
proposal distribution is known. Hence, the Metropolis-Hastings
algorithm can be implemented to obtain a sample from the posterior
distribution of $\bm z_c$. The marginal distribuions of the elements
of $\bm z_c$ can be easily obtained as well.

Regarding the marginals of $z_{-c,k}$, it is worth noting that at step
$j$ of the Metropolis-Hastings algorithm a conditional marginal
distribution on $\bm z_{c}^{(j)}$ (and the data $\bm y$) is obtained:
$\pi(z_{-c,k}| \bm z_{c}^{(j)}, \bm y)$. The posterior marginal can be
approximated by integrating over $\bm z_{c}$ as follows:
\begin{equation}
    \pi(z_{-c,k}| \bm y) = \int \pi(z_{-c,k}|
    \bm z_{c}, \bm y) \pi(\bm z_{c}|\bm y)
    d \bm z_{c}
    \simeq
    \frac{1}{N}
    \sum_{j=1}^N \pi(z_{-c,k}| \bm z_{c}^{(j)}, \bm y),
    \label{eq:marginal}
\end{equation}
where $N$ is the number of samples of the posterior distribution of
$\bm z_c$. That is, the posterior marginal of $z_{-c,k}$ can be
obtained by averaging the conditional marginals obtained at each
iteration of the Metropolis-Hastings algorithm.

\subsection{Effect of approximating the marginal likelihood}

So far, we have ignored the fact that the conditional marginal
likelihood $\pi(\bm y | \bm z_c)$ used in the acceptance probability
$\alpha$ is actually an approximation. In this section, we will
discuss how this approximation will impact the validity of the
inference.

The situations where a Metropolis-Hastings algorithm has inexact
acceptance probabilities are often called pseudo-marginal MCMC
algorithms and were first introduced in \citet{Beaumont:2003} in the
context of statistical genetics where the likelihood in the acceptance
probability is approximated using importance sampling.
\citet{AndrieuRoberts:2009} provided a more general justification of
the pseudo-marginal MCMC algorithm, whose properties are further
studied in \citet{Sherlocketal:2015} and
\citet{MedinaAguayoetal:2016}. These results show that if the (random)
acceptance probability is unbiased then the Markov chain will still
have as stationary distribution the posterior distribution of the
model parameters.

In our case, the error in the acceptiance rate is coming from a
deterministic estimate of the conditional marginal likelihood, hence
the framework of pseudo-marginal MCMC does not apply. However, since
it is deterministic, our MCMC chain will converge to \emph{a}
stationary distribution. This limiting distribution will be
\begin{equation}\label{eq1}%
    \tilde{\pi}(\bm z_c | \bm y) \propto \pi(\bm z_c)
    \tilde{\pi}(\bm y | \bm z_c)
\end{equation}
where the ``$\sim$'' indicates an approximation. \pkg{R-INLA} returns
an approximation to the conditional marginal likelihood term, which
implies an approximation to $\pi(\bm z_c | \bm y)$. This leaves the
question, about how good this approximation is, for which we have to
rely on asymptotic results, heuristics and numerical experience.

The conditional marginal likelihood estimate returned from
\pkg{R-INLA} is based on numerical integration and uses a sequence of
Laplace approximations \citep{isi:000264374200002,art632}. This
estimate is more accurate than the classical estimate using one
Laplace approximation. This approximation has, with classical
assumptions, relative error ${\mathcal O}(n^{-1})$ \citep{art367},
where $n$ is the number of replications in the observations. For our
purpose, this error estimate is sufficient, as it demonstrates that
\begin{equation}
    \label{eq:mlikerr}
    \frac{\tilde{\pi}(\bm z_c | \bm y)}{{\pi}(\bm z_c | \bm y)}
    \propto
    \frac{\tilde{\pi}(\bm y | \bm z_c)}{{\pi}(\bm y| \bm z_c)} =
    1 + {\mathcal O}(n^{-1})
\end{equation}
for plausible values of $\bm z_c$. However, as discussed by
\citet{isi:000264374200002,art632}, the classical assumptions are
rarely met in practice due to ``random effects'', smoothing etc.
Precise error estimates under realistic assumptions are difficult to
obtain; see \citet{art632} for a more detailed discussion of this
issue.

About numerical experience with the conditional marginal likelihood
estimate, \citet{HubinStorvik:2016b} have studied empirically its
properties and accuracy for a wide range of latent Gaussian models.
They have compared the estimates with those obtained using MCMC, and
in all their cases the approximates of the marginal likelihood
provided by INLA were very accurate. For this reason, we believe that
the approximate stationary distribution $\tilde{\pi}(\bm z_c | \bm y)$
should be close to the true one, without being able to quantify this
error in more details.

Although the error in Equation~(\ref{eq:mlikerr}) is pointwise, we do expect
the error would be smooth in $\bm z_c$. This is particularly
important, as in most cases we are interested in the univariate
marginals of $\tilde{\pi}(\bm z_c | \bm y)$. These marginals will
typically have less error as the influence of the approximation error
will be averaged out integrating out all the other components. A final
renormalization would also remove constant offset in the error.

Additionally, we will validate the approximation error in a simulation
study in Section~\ref{sec:Example} where we fit various models using
INLA, MCMC and INLA within MCMC and very similar posterior
distributions are obtained. Furthermore, the real applications in
Section~\ref{sec:Applications} also support that the approximations to
the marginal likelihood are accurate.

\subsection{Some remarks}

Common sense is still not out of fashion, hence there is an implicit
assumption that our INLA within MCMC approach should be only for models for
which it is reasonable to use the INLA-approach to do the inference
for the conditional model. The procedure that we have just shown will
allow INLA to be used together with the Metropolis-Hastings algorithm
(and, possibly, other MCMC methods) to obtain the posterior
distribution (and marginals) of $\bm z_c$ and the posterior marginals
of the elements in $\bm z_{-c}$. Hence, this will allow INLA to be
used to fit models not implemented in the \pkg{R-INLA} package as well
as providing other options for model fitting, that we summarise here.

The Metropolis-Hastings algorithm will allow any choice of the priors
on the set of parameters $\bm z_c$. This is an advantage (as shown in
the example in Section~\ref{subsec:lasso}) of combining MCMC and INLA
because priors that are not implemented in \pkg{R-INLA} can be used in
the model. In particular, improper flat priors, multivariate priors
and objective priors can be used.

The framework of conditional LGMs that we now can fit using our new
approach is quite rich. It includes models with missing covariates
that are imputed at each step of the Metroplis-Hastings algorithm (see
example in Section~\ref{subsec:miscov}), models with complex
non-linear effects in the linear predictor (see example in
Section~\ref{subsec:eco}) or models that have a mixture of effects in
the linear predictor \citep{Bivandetal:2015}.

\section{Simulation study}
\label{sec:Example}

In this section we develop simple examples to illustrate the method
proposed in the previous sections, and we investigate how this new
approach works in practice.

\subsection{Bivariate linear regression}
\label{subsec:ex:multi}

The first example is based on a linear regression with two covariates.
Our aim is to use our proposed method to obtain the posterior
distribution of the coefficients of the two covariates and then
compare the estimated marginals to the results obtained when the full
model is fitted with MCMC and INLA.

The simulated dataset contains 100 observations of a response variable
$\bm y$ and covariates $\bm u_1$ and $\bm u_2$. The model used to
generate the data is a typical linear regression, i.e.,

\begin{equation}
    y_i = \alpha + \beta_1 u_{1i} + \beta_2 u_{2i} +
    \varepsilon_i;\ i = 1, \ldots, 100
\end{equation}
Here, $\varepsilon_i$ is a Gaussian error term with zero mean and
precision $\tau$. The dataset has been simulated using $\alpha = 3$,
$\beta_1 = 2$, $\beta_2 = -2$ and $\tau = 1$. Covariates $u_{1i}$ and
$u_{2i}$ have also been simulated using a uniform distribution between
0 and 1 in both cases.

This model can be easily fitted using \pkg{R-INLA}. Given that we are
using a Gaussian model, inference is exact in this case (up to
integration error). For this reason, we can compare the marginal
distribution provided of $\beta_1$ and $\beta_2$ by INLA to the ones
obtained with our combined approach. Note that the Metropolis-Hastings
algorithm will provide the joint posterior distribution of
$\bm \beta = (\beta_1, \beta_2)$ that can be use to obtain the
posterior marginals of $\beta_1$ and $\beta_2$. Furthermore, we can
also compare the marginals of $\alpha$ and $\tau$, that will be
estimated by averaging the different conditional marginals obtained in
the Metropolis-Hastings steps.

In order to implement the Metropolis-Hastings algorithm to obtain a
sample from $\pi(\bm\beta|\bm y)$ we have chosen a starting point of
$\bm{\beta}^{(0)} = (0, 0)$. The transition kernel to obtain a
candidate $\bm{\beta}^{(t+1)}$ at iteration $t$ has been a bivariate
Guassian kernel centered at $\bm{\beta}^{(t)}$ with diagonal
variance-covariance matrix with values $1/0.75^2$ in the diagonal as
this provided a resonable acceptance rate. The prior distribution of
$\bm\beta$ has been the product of two Gaussian distributions with
zero mean and precission $0.001$ because these are the default priors
for linear effects in \pkg{R-INLA}.

\begin{figure}[h!]
    \begin{center}
        \includegraphics[width=14cm]{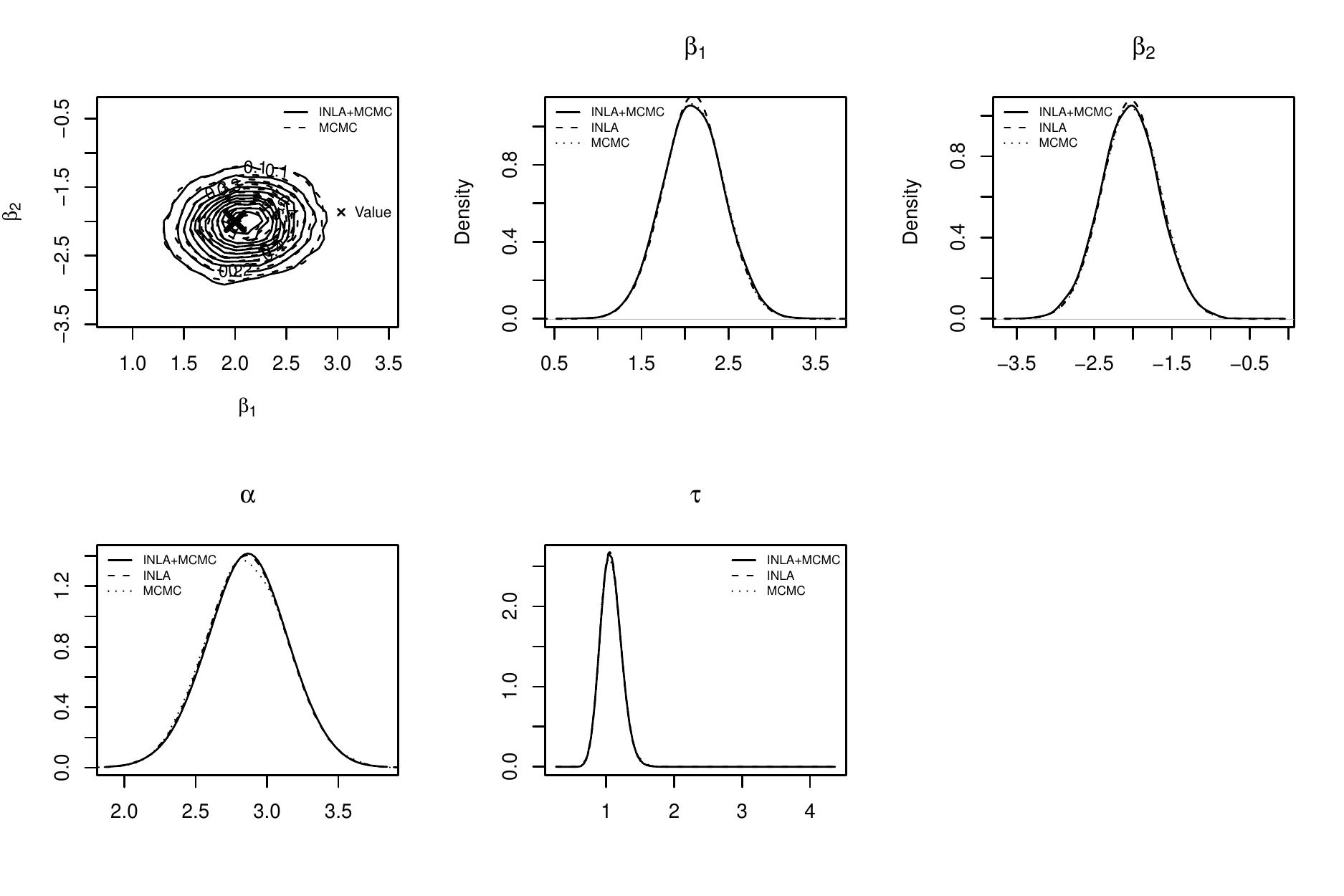}
    \end{center}
    \caption{Summary of results of model fitting combining INLA and
        MCMC in the bivariate case. Joint posterior distribution of
        $(\beta_1, \beta_2)$ (top-left plot) and posterior marginals
        of the model parameters. }
    \label{fig:toyexamplebiv}
\end{figure}

Figure~\ref{fig:toyexamplebiv} shows a summary of the results. Given
that both covariates are independent, their coefficients should show
small correlation and this can clearly seen in the plot of the joint
posterior distribution of $\bm\beta$. Also, it can be seen how the
marginals obtained with INLA within MCMC for $\beta_1$ and $\beta_2$
match those obtained with INLA and MCMC. In addition, we have included
the estimates of the posterior marginals of the intercept $\alpha$ and
the precission $\tau$. When using INLA within MCMC these are obtained
by Bayesian model averaging over the fitted models at every step of
the Metropolis-Hastings algorithm, whilst when computed with \pkg{R-INLA} these
are obtained by using INLA alone. The three estimation methods provide very
similar posterior distributions of the posterior marginals of the intercept and
the precission, which again confirms the accuracy of INLA within MCMC.

\subsection{Missing covariates}

In the next example, we will discuss the case of missing covariates.
In this example we will consider a linear regression with covariate
$\bm u_1$ only and we will assume that a number of values of the
covariates are missing. The aim is to include the imputation of this
variables into the model, so that the output is a marginal
distribution of the missing values. We will not discuss here the
different frameworks under which the values have gone missing, but
this is something that should take into account in the model. In
particular, we have removed the values of 9 covariates, which is
almost 10\% of our data and summary plots can be included in a 3x3
plot. Hence, in this case the missingness mechanism is of the type
missing completely at random \citep{LittleRubin:2002}.

Now, we will treat the missing values as if they were covariates. We
will use a block updating scheme as we can have a large number of
missing covariates. The transition kernel will be a multivariate
Gaussian with diagonal variance-covariance. The mean and variance for
all values are the mean and variance of the observed covariates,
respectively. The prior distribution is also a multivariate Gaussian,
but now with zero mean and diagonal variance-covariance matrix with
entries four times the variance of a uniform random variable in the
unit interval (the one used to simulate the covariates). This is done
so that the prior information is small compared to the information
provided by the covariates.

\begin{figure}[h!]
    \begin{center}
        \includegraphics[scale=.5]{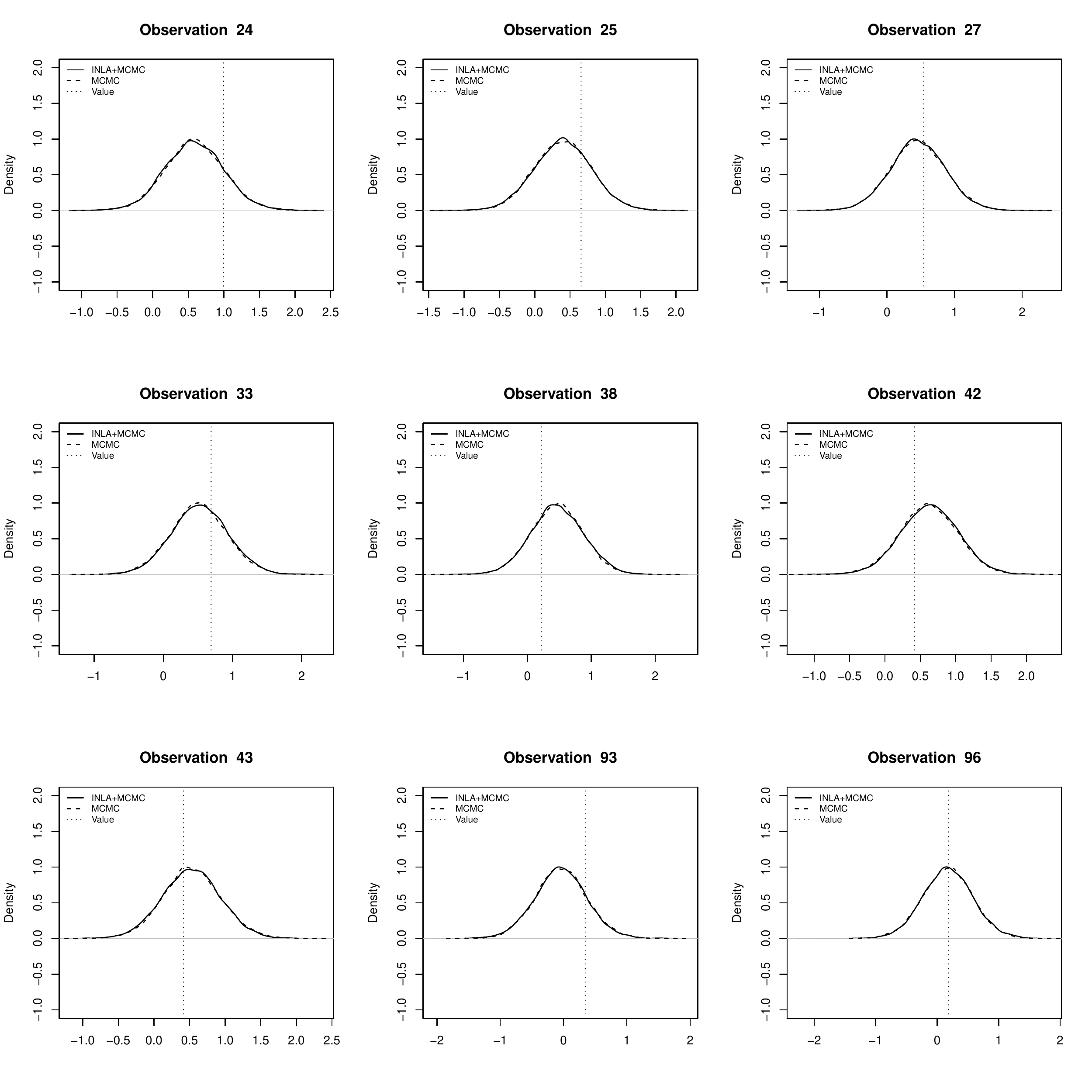}
    \end{center}
    \caption{Posterior marginals of the missing values in the
        covariates obtained by fitting a model with INLA within MCMC, and MCMC.}
    \label{fig:toyexamplemiscov}
\end{figure}

Figure~\ref{fig:toyexamplemiscov} shows the posterior marginals
obtained from the samples. As it can be seen, most of them are
centered at the actual values removed from the model. 
Note that this time the model with missing covariates cannot be fitted
with \pkg{R-INLA} so that we can only compare the marginals to those
obtained with MCMC. In all cases the marginals obtained with INLA
within MCMC and full MCMC are very similar.

\subsection{Poisson regression}

In this example we consider a Poisson regression with two covariates:
\begin{equation}
    y_i \sim Po(\mu_i);\ \log(\mu_i) = \alpha + \beta_1 u_{1i} + \beta_2 u_{2i};\
    i = 1, \ldots, 100.
\end{equation}
\noindent
The values of the parameters used to simulate the dataset are
$\alpha = 0.5$, $\beta_1 = 2$ and $\beta_2 = -2$.

As in Section~\ref{subsec:ex:multi}, our purpose is to estimate the
joint posterior distribution of $(\beta_1, \beta_2)$.
The prior distribution used now is the
same as in the previous example. Hence, the posterior marginal of
$\alpha$ is obtained by combining the different conditional marginals
obtained at the different steps of the Metropolis-Hastings algorithm.

\begin{figure}
    \centering
    \includegraphics[width=14cm]{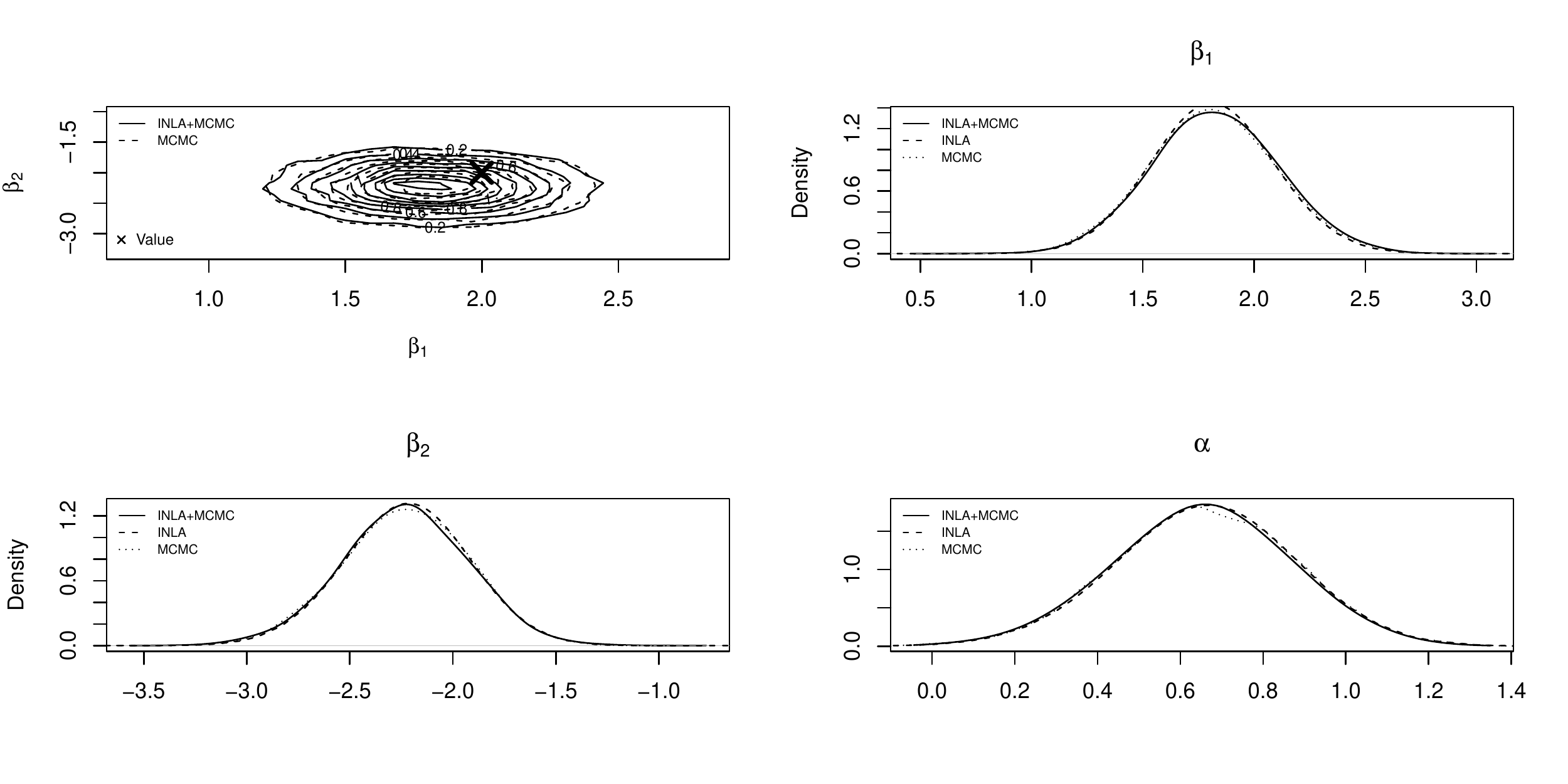}
    \caption{Summary of results of model fitting combining INLA and
        MCMC for the Poisson regression example. Joint posterior
        distribution of $(\beta_1, \beta_2)$ (top-left plot) and
        posterior marginals of the model parameters.}
    \label{fig:ex:Poisson}
\end{figure}

Figure~\ref{fig:ex:Poisson} shows the estimates of the marginal
distributions of the three parameters in the model, together with the
joint posterior distribution of $\beta_1$ and $\beta_2$. In all cases,
there is a very good agreement between the estimates obtained with
INLA, MCMC and INLA within MCMC of the marginals of the parameters in
the model.

\section{Applications}
\label{sec:Applications}

In this section, we will focus on some real life applications that provide a
more reallistic test of this methodology.  In all the examples, we have run
INLA within MC and MCMC for a total of 100500 simulations and discarded the
first 500.  Then we applied a thinning to keep one in ten iterations, to obtain
a final chain of 10000 samples. This includes samples from the missing
observations and fitted models. To fit the model using MCMC alone, we have used
\pkg{rjags} \citep{rjags:2016} with the same number of iterations and thinning. 

\subsection{Bayesian Lasso}
\label{subsec:lasso}

The Lasso \citep{Tibshirani:1996} is a popular regression and variable
selection method for variable selection. It has the nice property of
providing coefficient estimates that are exactly zero and, hence, it
performs model fitting and variable selection at the same time. For a
linear model with a Gaussian likelihood, the Lasso is trying to
estimate the regression coeffcient by minimising
$$
\sum_{i=1}^n \left( y_i - \beta_0 - \sum_{j=1}^p \beta_j
  x_{ij}\right)^2 + \lambda \sum_{j=1}^p |\beta_j|
$$
Here, $y_i$ is the response variable and $x_{ij}$ are associated
covariates. $n$ is the number of observations and $p$ the number of
covariates. $\lambda$ is a non-negative penalty term to control how
the shrinkage of the coefficients is done. If $\lambda = 0$ then the
fitted coefficients are those obtained by maximum likelihood, whilst
higher values of $\lambda$ will shrink the estimates towards zero.

The Lasso is closely related to Bayesian inference as it can be
regarded as a standard regression model with Laplace priors on the
variable coefficients. The Laplace distribution is defined as
$$
f(\beta) = \frac{1}{2\sigma}\exp\left(-\frac{|\beta -
      \mu|}{\sigma}\right),\ x\in \mathbb{R}
$$
where $\mu$ and $\sigma$, a positive number, are parameters of
location and scale, respectively. The Laplace prior distribution is
not available for (parts of) the latent field in \pkg{R-INLA}.
However, conditioning on the values of the $\bm \beta$-coefficients
the model can be easily fitted with \pkg{R-INLA}.

We will apply the methodology described in this paper to implement the
Bayesian Lasso by combining INLA and MCMC. We will be using the
\code{Hitters} dataset described in \citet{Jamesetal:2013}. This
dataset records several statistics about players in the Major League
Baseball, including salary in 1987, number of times at bat in 1986 and
other variables. Our aim is to build a model to predict the player's
salary in 1987 on some of the other variables recorded in 1986 (the
previous season).

We will focus on a smaller model than the one described in
\citet{Jamesetal:2013} and will consider predicting salary in 1987 on
only five variables measured from the 1986 season: number of times at
bat (AtBat), number of hits (Hits), the number of home runs (HmRun),
number of runs (Runs) and the number of runs batted (RBI).

For our implementation of the Bayesian Lasso, we
will be fitting models conditioning on the covariate coefficients
$\bm\beta = (\beta_1,\ldots\beta_p)$. Also, we will assume that
$\bm\beta$ and the error term precision $\tau$ are independent a
priori, i.e., $\pi(\bm\beta, \tau) = \pi(\bm\beta) \pi(\tau)$. This
will provide a simpler way to compare our results with the Lasso and
it will also make computations a bit simpler. However, note that
choosing it is also possible to choose a prior so that
$\pi(\bm\beta, \tau) = \pi(\bm\beta|\tau)\pi(\tau)$ \citep[see, for
example,][]{LykouNtzoufras:2011} The posterior distribution of these
variables will be obtained using MCMC.

\begin{figure}[h!]
    \begin{center}
        \includegraphics[width = 12cm]{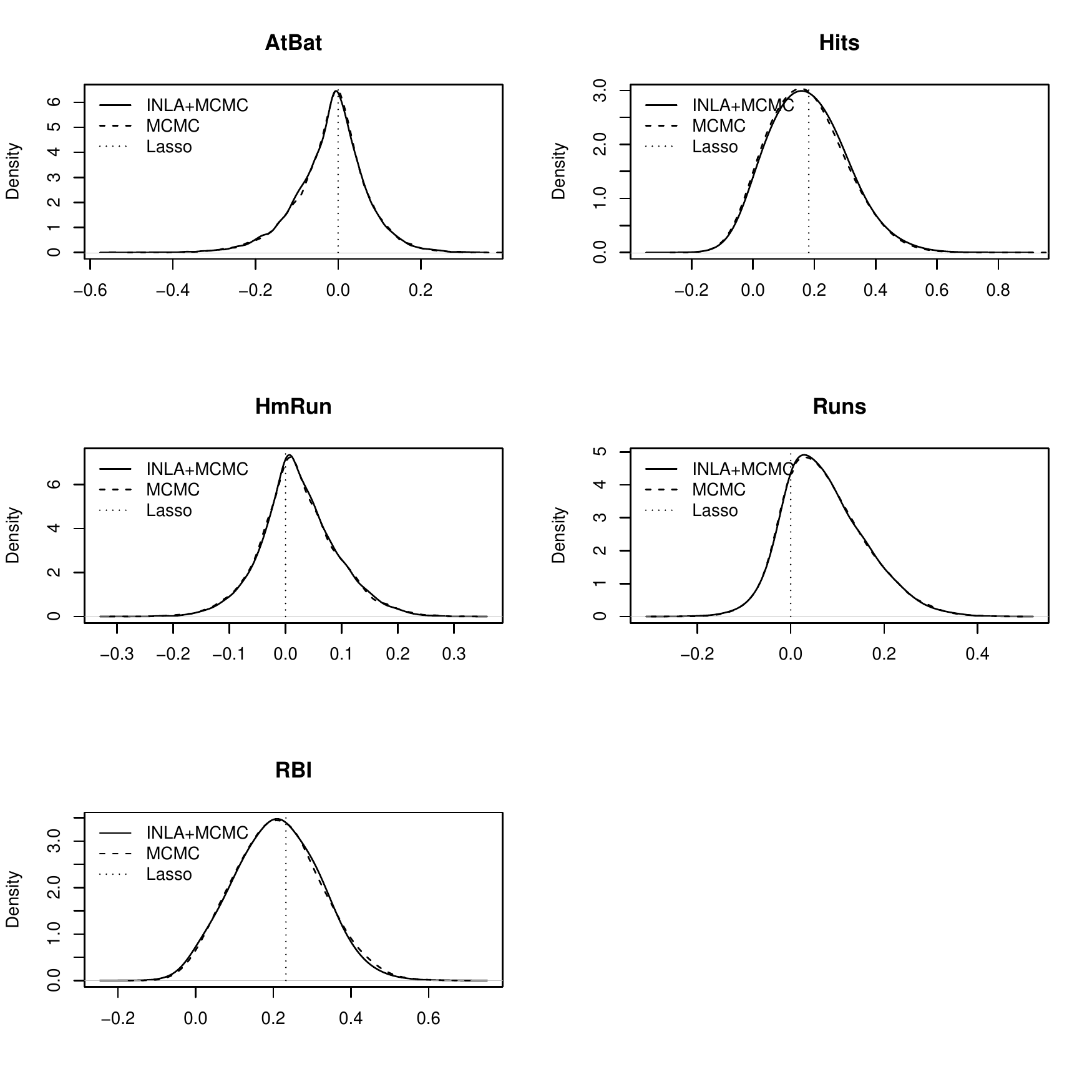}
    \end{center}
    \caption{Summary of results for the Lasso and Bayesian Lasso.}
    \label{fig:lasso}
\end{figure}

The summary of the Lasso estimates are available in Table
\ref{tab:lasso} and the posterior distribution of the coefficients is
in Figure \ref{fig:lasso}. In all cases, there is agreement between the
Lasso and Bayesian Lasso estimates. Also, the posterior distribution
of the model coefficients is the same for MCMC and combining INLA with
MCMC. For those coefficients with a zero estimate with the Lasso, the
posterior distribution obtained with the Bayesian Lasso is centered at
zero.
\begin{table}
    \centering
    \begin{tabular}{c|ccc}
      Coefficient & Lasso & INLA+MCMC & MCMC\\
      \hline
      AtBat & 0.00 & -0.01 (0.08) & -0.02 (0.08)\\
      Hits & 0.18 & 0.17 (0.12) & 0.17 (0.12)\\
      HmRun & 0.00 & 0.02 (0.07 )& 0.02 (0.07)\\
      Runs & 0.00 & 0.07 (0.09) & 0.07 (0.08)\\
      RBI & 0.23 & 0.21 (0.11) & 0.22 (0.11)\\
    \end{tabular}
    \caption{Summary estimates of the Lasso and Bayesian Lasso (posterior mean
        and standard deviation, between parenthesis).}
    \label{tab:lasso}
\end{table}

\subsection{Imputation of Missing Covariates}
\label{subsec:miscov}

\citet{JSSv045i03} describe the \RR package \pkg{mice} that implements
several multiple imputation methods. We will be using the
\code{nhanes} dataset to illustrate how our approach can be used to
provide imputation of missing coariates in a real dataset. This
dataset contains data from \citet{Schaffer:1997} on age, body mass
index (\code{bmi}), hypertension status (\code{hyp}) and cholesterol
level (\code{chl}). Age is divided into three groups: $20$-$39$,
$40$-$59$, $60$+.

Our aim is to impute missing covariates in order to fit a model that
explains the cholesterol level on age and body mass index. Although
the values of age have been completely observed, there are missing
values in body mass index and cholesterol level. INLA can handle
missing values in the response (and will provide a predictive
distribution) but, as already stated, is not able to handle models
with missing values in the covariates.

We will consider a very simple imputation mechanism by assigning a
Gaussian prior to the missing values of body mass index. This Gaussian
is centred at the average value of the observed values (26.56) and
variance four times the variance of the observed values (71.07,
altogether). With this, we expect to provide some guidance on how the
imputed values should be but allowing for a wide range of variation.
More complex imputation mechanisms could be considered \citep[see, for
example,][]{LittleRubin:2002}. As in previous examples, we will fit
the same model using MCMC in order to compare both results. The model
that we will fit is:
\begin{equation}
    \begin{array}{ccl}
      \textrm{chl}_i & = &\beta_0 + \beta_1 \textrm{bmi}_i +
                           \beta_2 \textrm{age2}_i+ \beta_3 \textrm{age3}_i + \varepsilon_i\\
      \beta_0 & \propto & 1\\
      \beta_k & \propto & N(0, 0.001);\ k = 1, 2, 3\\
      \varepsilon_i & \sim & N(0, \tau)\\
      \tau & \sim & Ga(1, 0.00005)
    \end{array}
    \label{eq:mi}
\end{equation}

\begin{figure}[h!]
    \centering
    \includegraphics[width = 10cm]{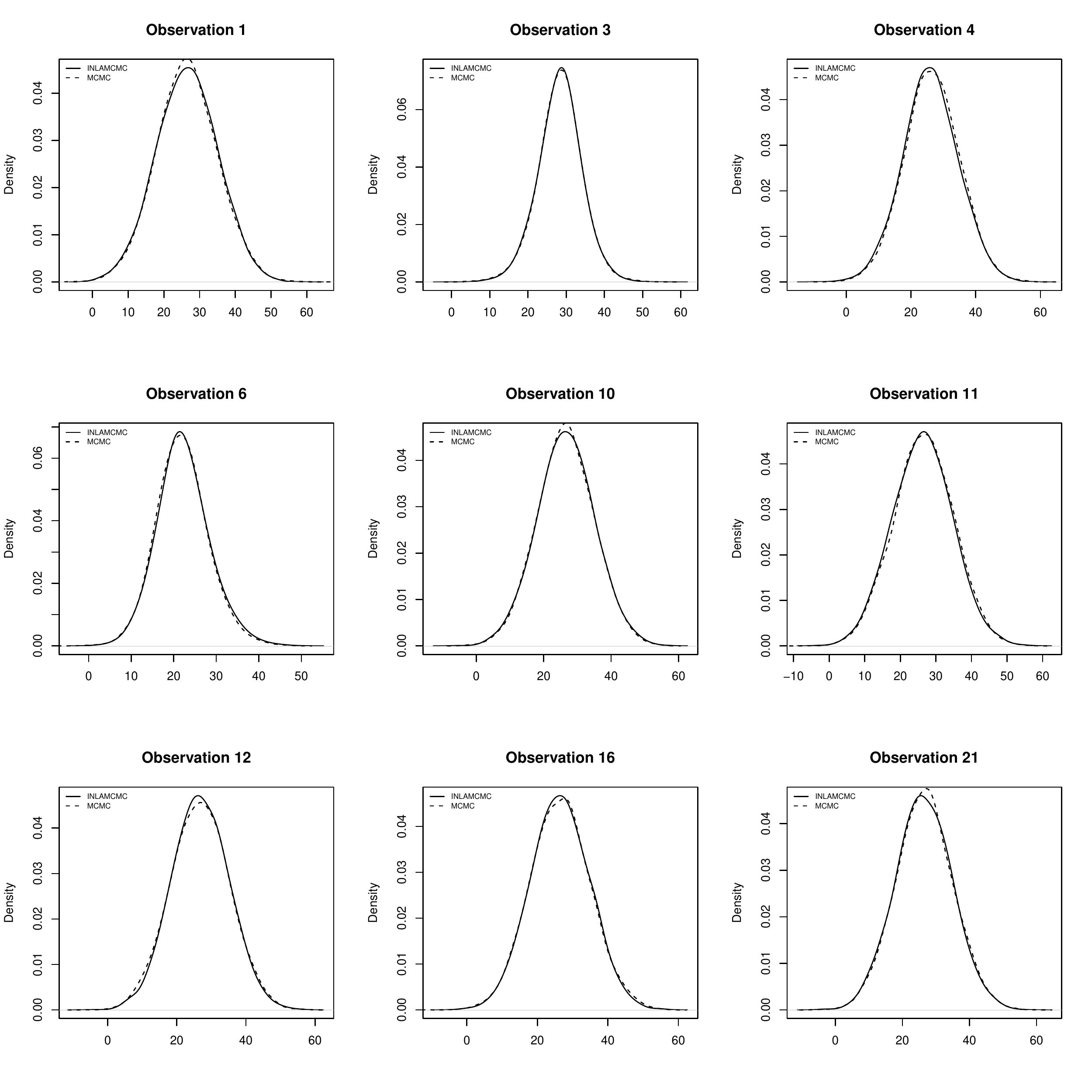}
    \caption{Marginal distributions of the imputed values of body mass
        index.}
    \label{fig:bmi}
\end{figure}

\begin{figure}[h!]
    \centering
    \includegraphics[width = 10cm]{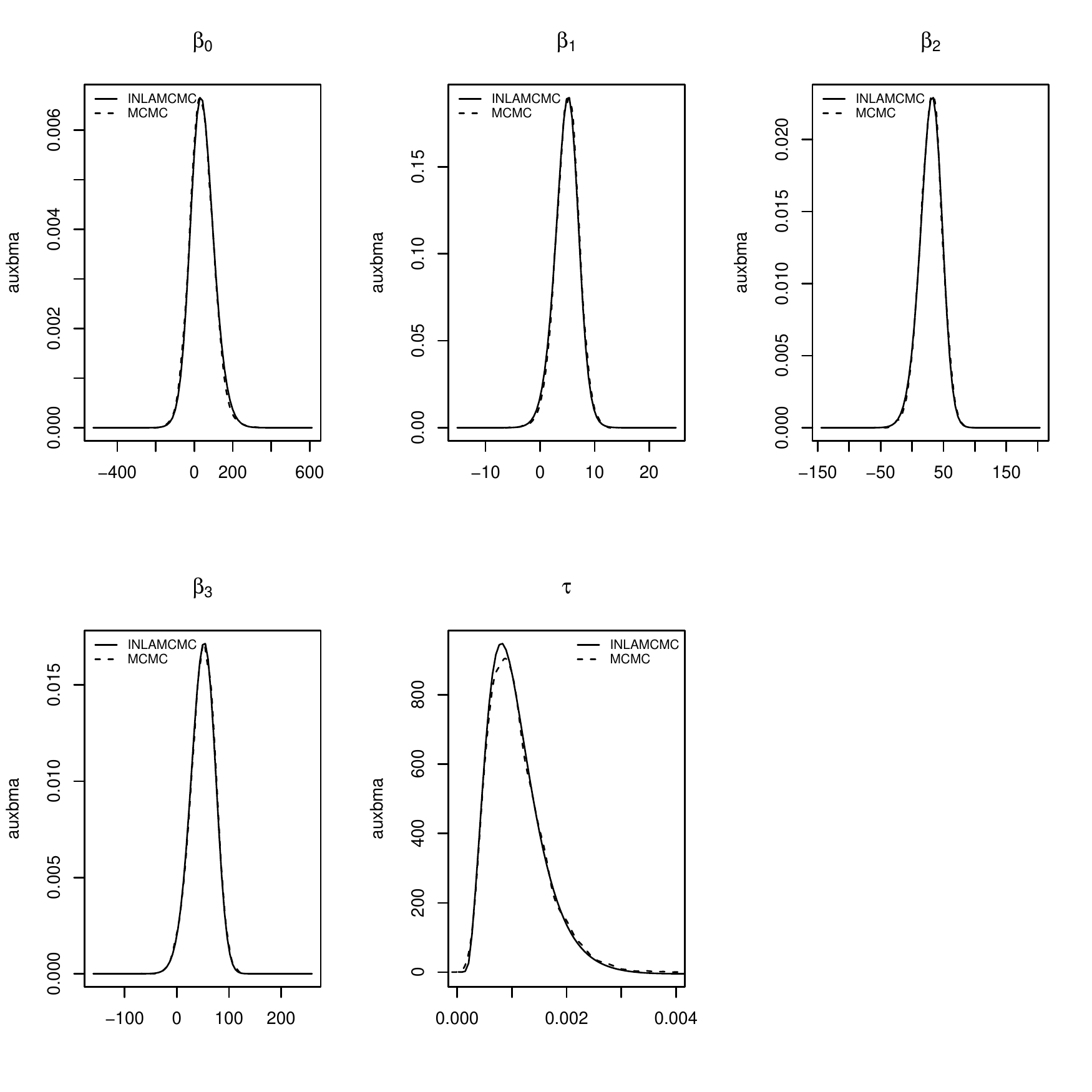}
    \caption{Marginal distributions of the model parameters, model
        with missing values in the covariates.}
    \label{fig:mi}
\end{figure}

\noindent
Figure \ref{fig:bmi} shows the posterior
marginal distributions of the imputed values of the body mass index.
Both MCMC and our approach provide very similar point estimates. Table
\ref{tab:bmi} summarises the model parameters obtained both with MCMC
and our approach and Figure \ref{fig:mi} displays the posterior
marginals of the model parameters obtained with our approach and MCMC.
In all cases, the marginals agree, and the point estimates look very
similar.

\begin{table}
    \centering
    \begin{tabular}{c|c|c}
      Parameter & MCMC & INLA+MCMC\\
      \hline
      $\beta_0$ & 39.760 (61.463) & 43.469 (62.603)\\
      $\beta_1$ & 4.994 (2.167) & 4.864 (2.206)\\
      $\beta_2$ & 29.989 (17.542)& 29.501 (17.871)\\
      $\beta_3$ & 50.049 (23.277) & 49.449 (23.207)\\
      \hline
      $\tau$ & 0.001 (0.0005) & 0.001 (0.0005)\\
    \end{tabular}

    \caption{Summary of model parameter posterior estimates: posterior mean
        and standard deviation (in parentheses), model with missing covariates.}
    \label{tab:bmi}
\end{table}

\subsection{Spatial econometrics models}
\label{subsec:eco}

\citet{Bivandetal:2014} describe a novel approach to extend the
classes of models that can be fitted with \pkg{R-INLA} to fit some
spatial econometrics models. In particular, they fit several
conditional models by fixing the values of some of the parameters in
the model, and then they combine these models using a Bayesian model
averaging approach \citep{Hoetingetal:1999}. \citet{Bivandetal:2015}
show a practical implementation with a spatial statistics model using
R package \pkg{INLABMA}. Some of these models have already been
included in \pkg{R-INLA} \citep{GomezRubioetal:2017} but are still
considered as experimental.

In this example we will focus on one of the spatial econometrics
models described in \citet{Bivandetal:2014} to illustrate how our new
approach to combine MCMC and \pkg{R-INLA} can be used to fit
unimplemented models. In particular, we will consider the spatial lag
model \citep{LeSagePace:2009}:
$$
\bm y = \rho \bm W \bm y + \bm X \bm \beta + \bm u;\ \bm u \sim N(0,
\sigma^2_u \bm I)
$$
Here, $\bm y$ is a vector of observations at $n$ areas, $\bm W$ is an
adjacency matrix, $\rho$ a spatial autocorrelation parameter, $\bm X$
a $n\times p$ matrix of covariates with associated coeffients
$\bm \beta = (\beta_1,\ldots,\beta_p)$ and $\bm u = (u_1,\ldots, u_n)$
an error term. $u_i,\ i=1,\ldots,n$, is Normally distributed with zero
mean and precision $\tau_u$. This model can be rewritten as follows:
$$
\bm y = (\bm I_n - \rho \bm W)^{-1}\bm X\bm\beta + \bm \varepsilon;\
\bm\varepsilon \sim N\left(0, \frac{1}{\tau_u} ((\bm I_n-\rho \bm W) (\bm
I_n-\rho \bm W^{\prime}))^{-1}\right)
$$
This model is difficult to fit with any standard software for
mixed-effects models because of parameter $\rho$. If the value of
$\rho$ is fixed, then it is easy to fit the model with \pkg{R-INLA} as
it becomes a linear term on the covariates plus a random effects term
with a known structure. Hence, by conditioning on the value of $\rho$
we will be able to fit the model with \pkg{R-INLA}. In order to use
our new approach, we will be drawing values of $\rho$ using MCMC and
conditioning on this parameter to fit the models with \pkg{R-INLA}.

Regarding prior distributions, $\rho$ is assigned a uniform between
$-1.5$ and $1$, $\beta_i,\ i = 1,\ldots, p$ a Gaussian prior with zero
mean and precision $0.001$ (the default), and $\tau_u$ is assigned a
Gamma distribution with parameters $1$ and $0.00005$ (the default for
the precision of a 'generic0' latent class in \pkg{R-INLA}).

We have fitted this model to the Columbus dataset available in \pkg{R}
package \pkg{spdep}. This dataset contains information about 49
neighbourhoods in Columbus (Ohio) and we have considered a model with
crime rates as the response and household income and housing value as
covariates. We have also fitted the spatial lag model using a maximum
likelihood approach, the method proposed by \citet{Bivandetal:2014}
and MCMC using an implementation of the model for the Jags software
included in package \pkg{SEjags}, which can be downloaded from Github.
The results are shown in Table \ref{tab:columbus}. All Bayesian
approaches have very similar estimates, and these are also very
similar to the maximum likelihood estimates.

\begin{table}
    \centering
    \begin{tabular}{crrrr}
      Parameter & MaxLik & INLA+MCMC & MCMC & INLA+BMA \\
      \hline
      Intercept & 61.05 (5.31) & 60.62 (6.08) & 58.53 (6.92) & 60.81 (5.33)\\
      $\beta_{\textrm{h. income}}$ & -1.00 (0.34) & -0.97 (0.37) & -0.91 (0.39) & -0.98 (0.33)\\
      $\beta_{\textrm{h. value}}$ & -0.31 (0.09) & -0.31 (0.09) & -0.30 (0.10) & -0.31 (0.09)\\
      $\rho$ & 0.52 (0.14) & 0.55 (0.13) & 0.55 (0.16) & 0.54 (0.11)\\
      $\tau_u$ & 0.01 (--) & 0.01 (0.002) & 0.01 (0.002) & 0.01 (0.00004)\\
      \hline
    \end{tabular}
    \caption{Posterior means (and standard deviation) of the spatial lag model
        fitted to the Columbus data set using three different methods.}
    \label{tab:columbus}
\end{table}

\section{Discussion}
\label{sec:Discussion}

In this paper, we have developed a novel approach to extend the models
that can be fitted with INLA. For this, we have used INLA within the
Metropolis-Hastings algorithm, so that only a few number of parameters
are sampled.

We have shown three important applications. In the first one, we have
implemented a Bayesian Lasso for variable selection using Laplace
priors on the coefficient of the covariates. By following this
example, other priors could be easily used with INLA. This includes
not only univariate priors but multivariate priors, that are seldom
available in \pkg{R-INLA}.

In our second example we have tackled the problem of imputation of
missing covariates in model fitting. Here, we have included a very
simple imputation method for the missing values in the covariates, so
that model fitting and imputation were done at the same time. Compared
to fitting the same model with MCMC, we obtained the same posterior
estimates. In an ongoing work, \citet{Camelettietal:2017} explore how
this can be extended to larger problems and how different imputation
models and missingness mechanisms can be properly addressed with INLA
and MCMC.

Finally, we have also shown how other models not included in the
\pkg{R-INLA} software can be fitted with INLA and MCMC. In particular,
we have fitted a spatial econometrics model by fitting conditional
models on the spatial autocorrelation parameter. This method can be
easily modified to suit any other models. In particular, Gibbs
sampling could be used if the full conditionals are available for a
subset of model parameters.

To sum up, we believe that this approach can be employed together with
INLA to fit more complex models and that it can also be combined with
other MCMC algorithms to develop simple samplers to fit complex
Bayesian hierarchical models. This method can work well when the
conditional models are hard to explore with current approaches for
which INLA provides a fast approximation, such as geostatistical
models. Furthermore, INLA could be embedded into a Reversible Jump
MCMC algorithm so that once the model dimension has been set, the
resulting model is approximated with INLA. See, for example,
\citet{Chenetal:2000} for a comprehensive list of MCMC algorithms that
could benefit from embedding INLA.

\section{Acknowledgements}

Virgilio G\'omez-Rubio has been supported by grant PPIC-2014-001, funded
by Consejer\'ia de Educaci\'on, Cultura y Deportes (JCCM) and FEDER,
and grant MTM2016-77501-P, funded by Ministerio de Econom\'ia y
Competitividad.

\bibliographystyle{Chicago} \bibliography{INLAMCMC}

\end{document}